\documentclass[%
superscriptaddress,showkeys,
12pt,
amsmath,amssymb,
aps,
]{revtex4-1}

\usepackage{graphicx}% Include figure files
\usepackage{dcolumn}% Align table columns on decimal point
\usepackage{bm}% bold math
\usepackage{xcolor}
\usepackage{url}
\usepackage{makecell}
\usepackage{subfigure}
\usepackage{multirow}
\usepackage{booktabs}
\usepackage{verbatim}
\usepackage{diagbox}

\newcommand{\moe}{\affiliation{Key Laboratory of Atomic and Subatomic Structure and Quantum Control (MOE), Guangdong Basic Research Center of Excellence for Structure and Fundamental Interactions of Matter, Institute of Quantum Matter, South China Normal University, Guangzhou 510006, China
}}

\newcommand{\iqm}{\affiliation{Guangdong-Hong Kong Joint Laboratory of Quantum Matter, Guangdong Provincial Key Laboratory of Nuclear Science, Southern Nuclear Science Computing Center, South China Normal University, Guangzhou 510006, China}}

\newcommand{\scnt}{\affiliation{Southern Center for Nuclear-Science Theory (SCNT), Institute of Modern Physics, Chinese Academy of Sciences, Huizhou 516000, Guangdong Province, China}}

\begin{document}
	
	%\preprint{APS/123-QED}
	
\title{Fifth Force and Hyperfine Splitting in Muonic Atoms}
	
\author{Jingxuan Chen}
%\email{jingxuanchen@impcas.ac.cn}
\moe\iqm
\affiliation{Institute of Modern Physics, Chinese Academy of Sciences, Lanzhou 730000, China}
\affiliation{University of Chinese Academy of Sciences, Beijing 100049, China}
	
\author{Qian Wu}
\affiliation{Institute of Modern Physics, Chinese Academy of Sciences, Lanzhou 730000, China}
	
\author{Xurong Chen}
\email{xchen@impcas.ac.cn}
\affiliation{Institute of Modern Physics, Chinese Academy of Sciences, Lanzhou 730000, China}
\affiliation{University of Chinese Academy of Sciences, Beijing 100049, China}
	
\author{Qian Wang}
\email{qianwang@m.scnu.edu.cn}
\moe\iqm\scnt 
	
\date{\today}% It is always \today, today,
	
\begin{abstract}

The potential existence of a fifth fundamental force, mediated by the X17 boson, has generated significant interest. This force can manifest itself as either a vector or pseudoscalar particle. In order to gain insight into the effective potentials produced by the X17 boson for hyperfine interactions in muonic systems, we conduct calculations for both the pseudoscalar and vector hypotheses. Our study reveals that, for muonic atoms with nuclear charge numbers ranging from 10 to 15, the energy shift associated with the fifth force can be as high as 0.61 eV. These effects have the potential to be detected through the utilization of high-precision X-ray detectors that are used to analyze the energy spectrum of muonic characteristic X-rays.
\end{abstract}
	
%Use showkeys class option if keyword
\maketitle
	
\section{Introduction}
\label{sec:intro}
	
Four fundamental forces - gravity, electromagnetism, strong and weak nuclear forces - have long been established as governing all interactions in the universe. The graviton, photon, gluon, and W and Z bosons mediate these forces, describing a range of phenomena. Recent explorations have questioned whether an additional force mediated by an unknown gauge boson exists. While supersymmetry and string theories suggest this possibility~\cite{Sushkov:2011md} ,experimental observations at the Institute for Nuclear Research of the Hungarian Academy of Sciences have presented a challenge to standard models~\cite{Krasznahorkay:2015iga,Krasznahorkay:2017gwn}, indicating the potential existence of a X17 particle (16.7 MeV) - mediated by a fifth force particle~\cite{Krasznahorkay:2015iga,Krasznahorkay:2017gwn,Krasznahorkay:2019lyl}- through a reaction$ { }^7 \mathrm{Li}+p \rightarrow{ }^8 \mathrm{Be}^* \rightarrow{ }^8 \mathrm{Be}+\gamma \rightarrow{ }^8 \mathrm{Be}+e^{+} e^{-} $.
	
This particle has been explained as a vector particle (dark photon with rest mass) and a pseudoscalar particle, both possessing the capability of coupling with light fermions and hadrons ~\cite{Feng:2016jff,Feng:2016ysn,Ellwanger:2016wfe}. It is important to note, however, that the findings of Refs. ~\cite{Feng:2016jff,Feng:2016ysn,Ellwanger:2016wfe} have yet to be independently verified and may have alternative interpretations and experimental searches. Despite this, we believe that consistent observations in two nuclear transitions warrant further investigation into the effects of the X17 boson in atomic systems. Specifically, we aim to determine the potential effects of a "light pseudoscalar Higgs"-type particle on atomic spectra. 
	
Although the particle can account for experimental observations, detecting it in atomic systems is challenging given its rest mass and the Compton wavelength.

Regrettably, the rest mass of the X17 particle in the observed range of 16.7 MeV makes it challenging to detect in atomic physics experiments. Moreover, the Compton wavelength of the X17 particle is smaller than the effective Bohr radius for both electron and muonic-bound systems, which makes it challenging to distinguish X17 effects from finite-nuclear-size effects in atomic spectroscopy experiments~\cite{Jentschura:2018zjv}. We propose that it may be more practical to examine muonic systems for evidence of the fifth force; due to the significantly smaller Bohr radius in comparison to electronic systems, they are far easier to be observed.
	
In fact, the fifth force effect presents in some extended "nuclear halo" due to its range of approximately $11.8 ~\mathrm{fm}$, which somewhat overlaps with the atomic nucleus. Additionally, the fifth force and the finite-nuclear-size are both proportional to a Dirac $\delta$ in the leading-order approximation, making it challenging to differentiate between them mathematically. To distinguish them, a system should be identified where the fifth force effect cannot be approximated using the Dirac-$\delta$ function. Instead, it can be described using a Yukawa potential\cite{Jentschura:2020zlr}. This means that two distinct mathematical models are required to accurately describe the fifth force and the finite-nuclear-size.
Furthermore, the definition of the Lamb shift~\cite{1990THEORY} explicitly excludes hyperfine structure which is utilized in many places~\cite{Jentschura:1996zz}. On the contrary, the hyperfine structure are induced by the Dirac-$\delta$ peak of the magnetic dipole field of the atomic nucleus at the origin in leading order for the S states. The leading order contribution to the hyperfine splitting is the Fermi contact interaction which is proportional to a Dirac-$\delta$ in coordinate space for the S states. This is consistent with the fact that the atomic nucleus has a radius not exceeding the femtometer scale which means the effect of short-range potentials would be less suppressed in the hyperfine splitting compared to the Lamb shift. 
    
To this end, we utilize the Gaussian expansion method~\cite{Hiyama2003} to derive the exact values of the fifth force interaction in both fine splitting and hyperfine splitting within muonic systems, rather than relying on ratios to other effects. We aim to distinguish the fifth force effect from the nuclear-size effects, as current experimental precision does not permit differentiation between the two. Our calculations are expected to yield more precise system wave functions, and we are also focused on exploring methods to detect and observe the effects of this fifth force. Encouragingly, a proof-of-concept experiment~\cite{PhysRevLett.130.173001} has demonstrated an exceptional level of precision, providing new hope for the discovery of the fifth force.
	
The present paper is structured as follows. Sec.~\ref{sec:Hamiltonian} presents the Hamiltonian of the interested system and related effective hyperfine interactions for both vector and pseudoscalar exchanges and the Gaussian expansion method utilized in our calculations. In Sec.~\ref{sec:result}, we report the results of our calculations, which include the precise numerical outcomes of the fifth force interaction for the S and P states and the anticipated value of the fifth force hyperfine structure for both vector and pseudoscalar particles. The conclusions and discussions of our study are outlined in Section~\ref{sec:summary}.

\section{The Methods }
\label{sec:Hamiltonian}
\subsection{The System Hamiltonian }

For a two-body system which consist of a muon and a nuclei, the Hamiltonian can be given as
\begin{equation}
    H_0=\frac{p^2}{2 \mu}+V,
\end{equation}
where $\mu$ is the reduce mass of the system, and $ V $ is the Coulomb potential.
	
The X17 particle induces a Yukawa potential
\begin{equation}
    \label{hf}
    H_F=\frac{h'_f h'_N}{4\pi r}e^{-m_X \, r} ,
\end{equation}
Here, we use the conventions that $h'_f = \varepsilon_f \, e$, $h'_N = \varepsilon_N \, e, e^2=4\pi\alpha$ for the coupling parameters of the hypothetical X17 vector particle~\cite{Jentschura:2020zlr} where the $\varepsilon^\prime_f$ and the $\varepsilon^\prime_N$ are the flavor-dependent coupling parameters of the fermions and the nucleons. In order to explain the 6.1 $\sigma$ peak observed in the experiments~\cite{Krasznahorkay:2015iga,Krasznahorkay:2017gwn}, the neutron coupling parameter is estimated to
\begin{equation}
    \left|\varepsilon_N\right|=\left|\varepsilon_u+2 \varepsilon_d\right| \approx\left|\frac{3}{2} \varepsilon_d\right| \approx \frac{1}{100}~.
\end{equation}
On the contrary, the proton has the relationship that $ \left|\varepsilon_N\right|=\left|2\varepsilon_u+ \varepsilon_d\right|\approx 0 $. This proves that the X17 vector particle is regarded as a “protophobic” particle. While the electronic coupling parameters is restricted by electron beam dump experiments which search for dark photons~\cite{Feng:2016ysn,Feng:2016jff} and the electron g $-$ 2 experiments set the high bound of the $\varepsilon_f$. Numerically, it fulfills the relationship
\begin{equation}
    2\times10^{-4} < \varepsilon_f < 1.4\times10^{-3}.
\end{equation}

Finite-nuclear-size(HFS) is necessary to distinguish from Yukawa potential in mathematical approximation forms.
This potential Hamiltonian includes a Dirac - $\delta$ potential, which is a good approximation for the short-range effect of the finite-nuclear-size. 
\begin{equation}
    \label{hfns}
    H_{\mathrm{FNS}}=\frac{2\pi}{3}Z\alpha r_n^2\delta^{(3)} \, (\vec{r}),
\end{equation}
where $r_n=\sqrt{\langle r_n^2\rangle}$ is the root-mean-square charge radius of the nucleus and its value is given by Ref.~\cite{Angeli:2004kvy}.By the calculation of Sec. \ref{sec:result}, we find it's difficult to separate the potential of the fifth force from the finite-nuclear-size while the range of the fifth force interaction somewhat overlaps with the atomic nucleus in fact.~\cite{Jentschura:2018zjv}
	
After that, we consider the impact of the hyperfine structure of the potential of the fifth force because the effect of the short range potentials is less suppressed in the hyperfine splitting.~\cite{Jentschura:2020zlr} When X17 boson is a vector particle, its Hamiltonian is given by
\begin{equation}
    \label{hhfsv}
    \begin{aligned}
	H_{\mathrm{HFS}, V}= & \frac{h_f^{\prime} h_N^{\prime}}{16 \pi m_f m_N}\left[-\frac{8 \pi}{3} \delta^{(3)}(\vec{r}) \vec{\sigma}_f \cdot \vec{\sigma}_N\right. \\
	& -\frac{m_X^2\left(\vec{\sigma}_f \cdot \vec{r} \vec{\sigma}_N \cdot \vec{r}-r^2 \vec{\sigma}_f \cdot \vec{\sigma}_N\right)}{r^3} \mathrm{e}^{-m_X r} \\
	& -\left(1+m_X r\right) \frac{3 \vec{\sigma}_f \cdot \vec{r} \vec{\sigma}_N \cdot \vec{r}-r^2 \vec{\sigma}_f \cdot \vec{\sigma}_N}{r^5} \mathrm{e}^{-m_X r} \\
	& \left.-\left(2+\frac{m_f}{m_N}\right)\left(1+m_X r\right) \frac{\vec{\sigma}_N \cdot \vec{L}}{r^3} \mathrm{e}^{-m_X r}\right] .
    \end{aligned}
\end{equation}
where we take the maximum value of $h_{f}^{\prime} = 1.4\times10^{-3} e$. 
	
Take a massless limit $ m_X \to 0 $ and make the replace that $h_{f}^{\prime}\to e$, $h_{N}^{\prime}\to -e$, the Yukawa potential recovers the Coulomb potential $H_0 \to - \frac{e^2}{4\pi\:r}=-\frac{\alpha}{r}$. In this way, the Fermi Hamiltonian $H_C$ could be recovered (the Eq.~(10) of Ref.~\cite{PhysRevA.73.062503}), through making replacements that 
 $h'_f\to e,~~h'_N\to\frac{\text{g}_N\left(-e\right)}{2}=\frac{\text{g}_N\left|e\right|}{2}$ ~\cite{Jentschura:2020zlr}.
\begin{equation}
    \label{hc}
    \begin{aligned}
    H_C=&\frac{\text{g}_N\alpha}{m_f\:m_N}\Bigg[\frac{\pi}{3}\:\vec{\sigma}_f\cdot\vec{\sigma}_N\:\delta^{(3)}(\vec{r})\\ &+\frac{3\:\vec{\sigma}_f\cdot\vec{r}\:\vec{\sigma}_N\cdot\vec{r}-r^2\:\vec{\sigma}_f\cdot\vec{\sigma}_N}{8\:r^5}+\frac{\vec{\sigma}_N\cdot\vec{L}}{4\:r^3} \Bigg]  .
    \end{aligned}
\end{equation}
For a pseudoscalar X17 particle, we have
    \begin{equation}
    \label{hhfsa}
    \begin{aligned}
    H_{\mathrm{HFS}, A}= & \frac{h_f h_N}{16 \pi m_f m_N}\left[\frac{4 \pi}{3} \delta^{(3)}(\vec{r}) \vec{\sigma}_f \cdot \vec{\sigma}_N\right. \\
    & -\frac{m_X^2 \vec{\sigma}_f \cdot \vec{r} \vec{\sigma}_N \cdot \vec{r}}{r^3} \mathrm{e}^{-m_X r}+\left(1+m_X r\right) \\
    & \left.\times \frac{3 \vec{\sigma}_f \cdot \vec{r} \vec{\sigma}_N \cdot \vec{r}-\vec{\sigma}_f \cdot \vec{\sigma}_N r^2}{r^5} \mathrm{e}^{-m_X r}\right] .
    \end{aligned}
\end{equation}
Here the pseudoscalar coupling parameters have been estimated as the followed function form ~\cite{Jentschura:2020zlr,Ellwanger:2016wfe,Cheng:2012qr} which is inspired by an analogy with putative pseudoscalar Higgs couplings~\cite{Cheng:2012qr}.
\begin{equation}
    h_f=\xi_f \frac{m_f}{v}, \quad h_N=\xi_N \frac{m_N}{v},
\end{equation}
where the coupling parameters $ v =246$ GeV is the vacuum expectation value of the Higgs field, ${m_f}$ is the fermion mass, ${m_N}$ is the nucleon’s mass. According to Ref.~\cite{Jentschura:2020zlr}, the coupling parameters could be estimated to $h_{N}=-2.3\times10^{-3}$ and $h_{f}=-3.8\times10^{-4}$.
	
The Hamiltonian is different in S states and P states because of the spin-dependence Hamiltonian. The following simplifications have been made for the sake of simplicity in the subsequent calculations.
	
For S states, 
\begin{equation}
\label{reduce}
    \vec{\sigma}_{f}\cdot\vec{r}~\vec{\sigma}_{N}\cdot\vec{r}~\to~\frac{1}{3}r^2\vec{\sigma}_f\cdot\vec{\sigma}_N,\quad \vec{\sigma}_N\cdot\vec{L}~\to~0.
\end{equation}  
So the Hamiltonian can be reduced to 
\begin{equation}
    \label{hhfsvs}
    H_{\mathrm{HFS},V}=-\frac{h'_fh'_N\vec\sigma_f\cdot\vec\sigma_N}{24\pi m_fm_N}\bigg[4\pi\delta^{(3)}(\vec r)-\frac{m_X^2}{r}\mathrm e^{- m_X \, r}\bigg], 
\end{equation}
\begin{equation}
    \label{hhfsas}
    H_{HFS,A}=\frac{h_fh_N\vec{\sigma}_f\cdot\vec{\sigma}_N}{48\pi m_f m_N}\bigg[4\pi\delta^{(3)}(\vec{r})-\frac{m_X^2}{r}\mathrm{e}^{-m_X \, r}\bigg],
\end{equation}
\begin{equation}
    \label{hcs}
    H_C=\frac{\pi\:\text{g}_N\alpha}{3\:m_f\:m_N}\:\vec{\sigma}_f\cdot\vec{\sigma}_N\:\delta^{(3)}(\vec{r}).	
\end{equation}
	
When we calculate the expected value of these Hamiltonian, the spin-independent terms is easy to calculate by the GEM but the spin-dependent terms calculate indirectly. And the expected value of spin terms $\left\langle \vec{\sigma}_f\cdot\vec{\sigma}_N\right\rangle $ is connected with the nuclear spin. Nuclear spin can be divided into three types: 1/2, 1 and 3/2 and the expected values are showed in Table~\ref{table:S-nuclear spin} below.
\begin{table*}[htbp]
    \caption{~The expected values of the nuclear spin term $\left\langle \vec{\sigma}_f\cdot\vec{\sigma}_N\right\rangle $ in $ S_{1/2}$ state. The $F$ represents the spin of the total system and the s represents the nuclear spin.}
    \setlength{\tabcolsep}{30pt}
    \centering
    \begin{tabular}{ c  c  c }
    \hline
    \hline
    & $ F = J - s $ &  $ F = J + s $  \\
    \hline
    s = 1/2 & -3 & 1 \\
    s = 1 & -4 & 2 \\
    s = 3/2 & 2 & $-\frac{3}{2}\sqrt{6}$ \\
    \hline
    \hline  
    \end{tabular} 
    \label{table:S-nuclear spin}
\end{table*}
	
For $P_{1/2}$ states, the contact potential is zero and the Hamiltonian can be reduced to
\begin{equation}
    \begin{aligned}
    H_{\mathrm{HFS}, V}= & \frac{h_f^{\prime} h_N^{\prime}}{16 \pi m_f m_N}\left[-\frac{m_X^2\left(\vec{\sigma}_f \cdot \vec{r} \vec{\sigma}_N \cdot \vec{r}-r^2 \vec{\sigma}_f \cdot \vec{\sigma}_N\right)}{r^3} \right. \\
    & \times\mathrm{e}^{-m_X r}-\left(1+m_X r\right) \frac{3 \vec{\sigma}_f \cdot \vec{r} \vec{\sigma}_N \cdot \vec{r}-r^2 \vec{\sigma}_f \cdot \vec{\sigma}_N}{r^5} \\
    & \times\mathrm{e}^{-m_X r}\left.-\left(2+\frac{m_f}{m_N}\right)\left(1+m_X r\right) \frac{\vec{\sigma}_N \cdot \vec{L}}{r^3} \mathrm{e}^{-m_X r}\right] ,
    \end{aligned}
\end{equation}

\begin{equation}
    \begin{aligned}
    H_{\mathrm{HFS}, A}= & \frac{h_f h_N}{16 \pi m_f m_N}\left[-\frac{m_X^2 \vec{\sigma}_f \cdot \vec{r} \vec{\sigma}_N \cdot \vec{r}}{r^3} \mathrm{e}^{-m_X r}\right. \\
    & \left.+\left(1+m_X r\right)\frac{3 \vec{\sigma}_f \cdot \vec{r} \vec{\sigma}_N \cdot \vec{r}-\vec{\sigma}_f \cdot \vec{\sigma}_N r^2}{r^5} \mathrm{e}^{-m_X r}\right] ,
    \end{aligned}
\end{equation}

\begin{equation}
    H_C=\frac{\mathrm{g}_N \alpha}{m_f m_N}\bigg[\frac{3 \vec{\sigma}_f \cdot \vec{r} \vec{\sigma}_N \cdot \vec{r}-r^2 \vec{\sigma}_f \cdot \vec{\sigma}_N}{8 r^5}+\frac{\vec{\sigma}_N \cdot \vec{L}}{4 r^3}\bigg].
\end{equation}
The Table~\ref{table:P-nuclear spin} below presents the expected values of the nuclear spin dependence term:
    \begin{table*}[htbp]
    	\caption{~The expected values of the nuclear spin term in $ P_{1/2}$ state.}
    	\setlength{\tabcolsep}{5pt}
    	\centering
    	\begin{tabular}{ c  |cc  |cc  |cc }
    		\hline
    		\hline
    		Nuclear spin &\multicolumn{2}{c|}{$ \langle\vec{\sigma}_f\cdot\vec{r}\vec{\sigma}_N\cdot\vec{r}\rangle $}& \multicolumn{2}{c|}{$ \langle\vec{\sigma}_N\cdot\vec{L}\rangle$} & \multicolumn{2}{c}{$\left\langle \vec{\sigma}_f\cdot\vec{\sigma}_N\right\rangle$} \\
    		& $ F = J - s $ &  $ F = J + s $  &$ F = J - s $&$ F = J + s $ & $ F = J - s $ &  $ F = J + s $ \\
    		
    		\hline
    		s = 1/2 & $(1+\frac{4\sqrt{10}}{15})r^2$ & $(-\frac{1}{3}-\frac{4\sqrt{10}}{45})r^2$ & $-2$ & $\frac{2}{3} $ & $1$ & $-\frac{1}{3}$\\
    		%\hline
    		s = 1 & $(-4+\frac{16\sqrt{10}}{45})r^2$ & $(2-\frac{8\sqrt{10}}{45})r^2$ & $-\frac{8}{3}$ & $\frac{4}{3}$ & $-4$ & $2$\\
    		%\hline
    		s = 3/2 & $(5-\frac{4\sqrt{10}}{9})r^2$ & $(-3+\frac{4\sqrt{10}}{15})r^2$ & $\frac{10}{3}$ & $ -2 $ & $ 5 $ & $-3$ \\
    		\hline
    		\hline
    	\end{tabular}
    	\label{table:P-nuclear spin}
    \end{table*}

Next, we will employ the Gaussian expansion method to calculate the impact of the fifth force on muonic atoms using these Hamiltonians.

\subsection{Gaussian expansion method}
\label{sec:Gaussian}
	
The two-body system can be described using the time-independent Schrödinger equation with the Coulomb potential $ V(r) $:
	\begin{equation}
		\left[ -\frac{\hbar^{2} } {2\mu} \nabla ^{2}+ V(r) -E \right] \psi_{l m}(\mathbf{r}) = 0
	\end{equation}
where $\mu$ represents the reduced mass of the system and $\psi_{l m}$ denotes the wavefunction with orbital quantum number $l$ and magnetic quantum number $m$. 
In the GEM, the wavefunction is expanded using a set of Gaussian basis functions\cite{Hiyama2003}
	\begin{equation}
		\begin{aligned}
			&\psi_{l m}(\mathbf{r})=\sum_{n=1}^{n_{\max }} c_{n l} \phi_{n l }^{\mathrm{G}}(\mathbf{r}) Y_{l m}(\mathbf{r}) , \\
			&\phi_{n l}^{\mathrm{G}}(r)=\left(\frac{2^{l+2}\left(2 \nu_n\right)^{l+\frac{3}{2}}}{\sqrt{\pi}(2 l+1) ! !}\right)^{\frac{1}{2}} r^l e^{-\nu_n r^2}.
		\end{aligned}
	\end{equation}
where the Gaussian range parameters are chosen to be in geometric progression:
    \begin{equation}
    	\begin{aligned}
    	&\nu_n=1/r_n^2 \\
    	r_{n}=r_{\rm min} &a^{n-1},  \quad (n=1 \sim n^{\rm max})
    \end{aligned}
    \end{equation}
The expansion coefficients ${c_{n l}}$ and the eigenenergy $E$ can be determined by the Rayleigh-Ritz variational principle, which results in a generalized matrix eigenvalue problem:
	\begin{equation}
		\sum_{n^{\prime}=1}^{n_{\max }}\left[T_{n n^{\prime}}+V_{n n^{\prime}}-E N_{n n^{\prime}}\right] c_{n^{\prime} l}=0, (n=1-n_{max}).
	\end{equation}
The three Gaussian parameters $ \{ n_{max},~ r_{1},~ r_{n_{max}} \}$ used in this paper have been fine-tuned and optimized to achieve the best results. Specifically, we have used $ n_{max}=90,~ r_{1}=0.001 \times r_{B}, ~ r_{n_{max}}=70 \times r_{B} $, where the $r_{B}$ represents the Bohr radius of system.

\section{Numerical results}
\label{sec:result}
	
\subsection{ The Effect of fine Structure of Fifth Force}%3.1
	
It is well-known that detecting the fifth force in electric systems is challenging. Additionally, due to the mass difference, the Bohr radius of muonic deuterium is approximately 200 times larger than that of regular deuterium. To verify this and the accuracy of the fifth force model, we calculate the fifth force in both deuterium and muonic deuterium. As a result, the fifth force in muonic deuterium is $ (m_{\mu}/m_{e})^{3} \approx 10^{7} $ stronger than that in deuterium. From the Table.\ref{table:fif}, it is evident that the fifth force has larger impact on the two-body system in the S states than that in the P states due to the differences in the wave function.
	
	\begin{table*}[htbp]
		\caption{~Fifth force calculation results in deuterium and muonic deuterium.}
		\setlength{\tabcolsep}{10pt}
		\centering
		\begin{tabular}{ c  c  c }
			\hline
			\hline
			system & deuterium & muonic deuterium \\
			\hline
			mass (MeV) & 0.511 & 105.658 \\
			%\hline
			Bohr radius ($ \times 10^3 $ fm) & 52.918 & 0.285 \\
			%\hline
			$ \Delta E_{F} (1S(1/2)) $ ($\mu$eV) & $ 1.0425 \times 10^{-6} \sim 7.2975 \times 10^{-6} $ & $ 6.6216 \sim 46.3514 $ \\
			%\hline
			$ \Delta E_{F} (2S(1/2)) $ ($\mu$eV) &  $ 1.3031 \times 10^{-7} \sim 9.1218 \times 10^{-7} $ & $ 0.8255 \sim 5.7792 $ \\
			%\hline 
			$ \Delta E_{F} (2P(1/2))$ ($\mu $eV)  & $ 3.2400 \times 10^{-15} \sim 2.2679 \times 10^{-14} $ & $ 7.8585 \times 10^{-4} \sim 5.5009 \times 10^{-3} $ \\
			\hline
			\hline
		\end{tabular}
		\label{table:fif}
	\end{table*}
	
\begin{figure}[htbp]
		\centering
		\includegraphics[scale=0.33]{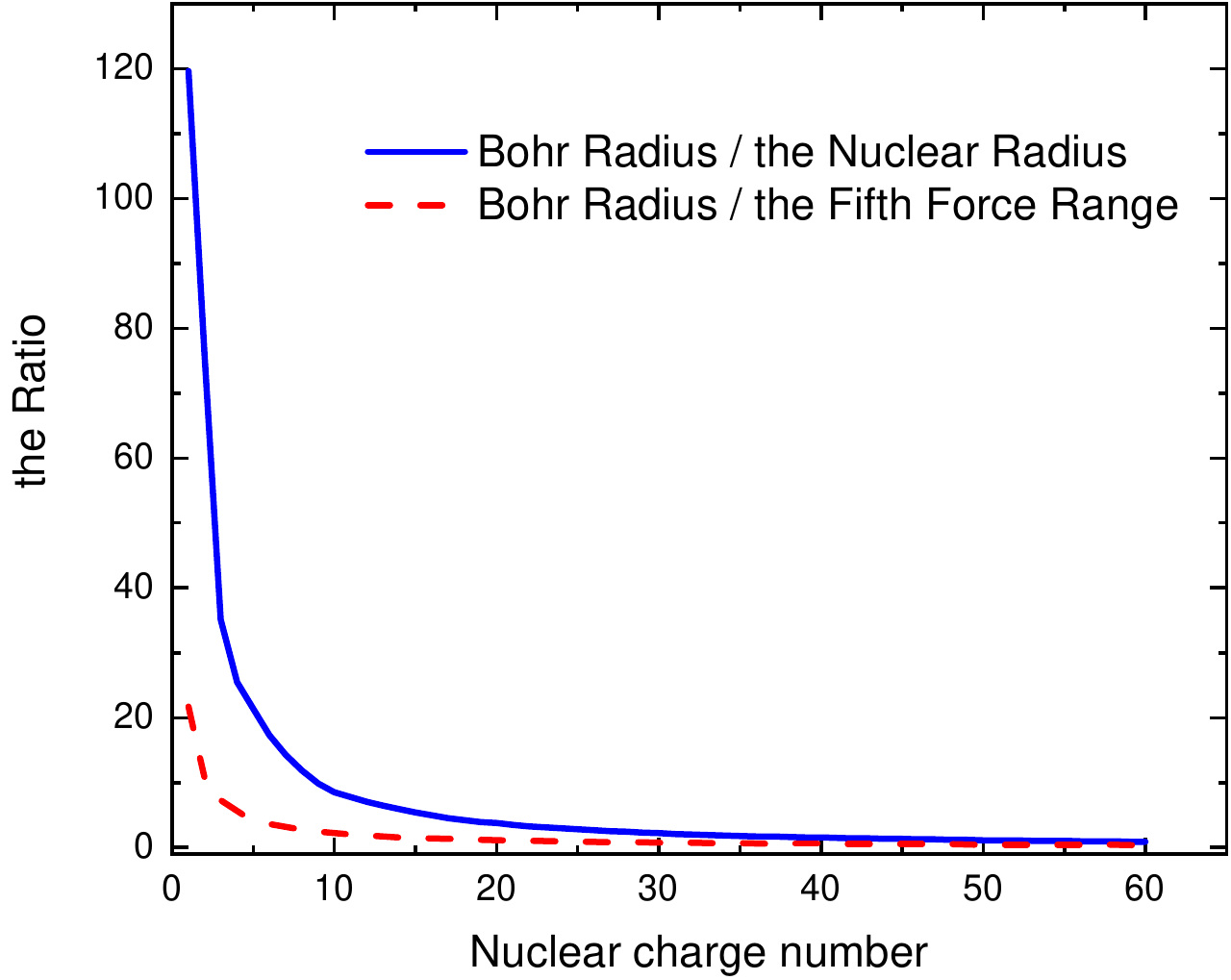}
		\caption{ Compare the ratio of Bohr radius to nuclear radius and the ratio of Bohr radius to the fifth force range to find the particle that fits the approximation. It is easy to find the Bohr radius of muonic atom with the nuclear charge number from 6 to 15 can satisfy the conditions well. The blue curve represents the ratio of Bohr radius to the nuclear radius. While the red dashed curve represents the ratio of Bohr radius to the range of the fifth force. } 
		\label{fig:Bohr}
	\end{figure}

In order to anaylze the effect of fifth force in S states, it is necessary to distinguish between the fifth force and finite-nuclear-size. The Dirac-$\delta$ approximate needs that the Bohr radius is much larger than the nuclear radius. At the same time,the Yukawa potential needs the Bohr radius is in the same order of magnitude as the nuclear radius. In Fig.~\ref{fig:Bohr}, we find muonic atoms with nuclear charge number from 6 to 15 can satisfy the approximate conditions. And Table.~\ref{table:Expected value} shows the expected value of them.
	
		\begin{table*}[htbp]
		\caption{~The fifth force effect of 1S, 2S and 2P states in muonic atom with nuclear charge number from 6 to 15.}
		\setlength{\tabcolsep}{10pt}
		\centering
		\begin{tabular}{ c  c  c  c  c  c }
			\hline
			\hline
			Nuclear charge number & 6 & 7 & 8 & 9 & 10 \\
			\hline
			 $ \Delta E_{F} (1S(1/2))$ (eV)  & $-0.0397$ & $-0.0562$ & $-0.0754$ & $-0.139$ & $-0.155$ \\
			%\hline
			$ \Delta E_{F} (2S(1/2)) $ (eV) & $-0.00468$ & $-0.00655$ & $-0.00867$ & $-0.0158$ & $-0.0174$ \\
			%\hline
			 $ \Delta E_{F} (2P(1/2)) $ (meV) & $-0.170$ & $-0.319$ & $-0.544$ & $-1.24$ & $-1.66$ \\
			\hline
			\hline
			Nuclear charge number & 11 & 12 & 13 & 14 & 15 \\
			\hline
			$ \Delta E_{F} (1S(1/2))$ (eV) & $-0.231$ & $-0.374$ & $-0.320$ & $-0.503$ & $-0.610 $\\
			%\hline
			$ \Delta E_{F} (2S(1/2))$ (eV) & $-0.0256$ & $-0.0411$ & $-0.0349$ & $-0.0544$ & $-0.0655$\\
			%\hline
			$ \Delta E_{F} (2P(1/2))$ (meV) & $-2.89$ & $-5.42$ & $-5.27$ & $-9.32$ & $-12.58$\\
			\hline
			\hline
		\end{tabular}
		\label{table:Expected value}
	\end{table*}
	
	\begin{figure*}[htbp]
		\centering
		\subfigure[]
		{
			\begin{minipage}[htbp]{.45\linewidth}
				\centering
				\includegraphics[scale=0.33]{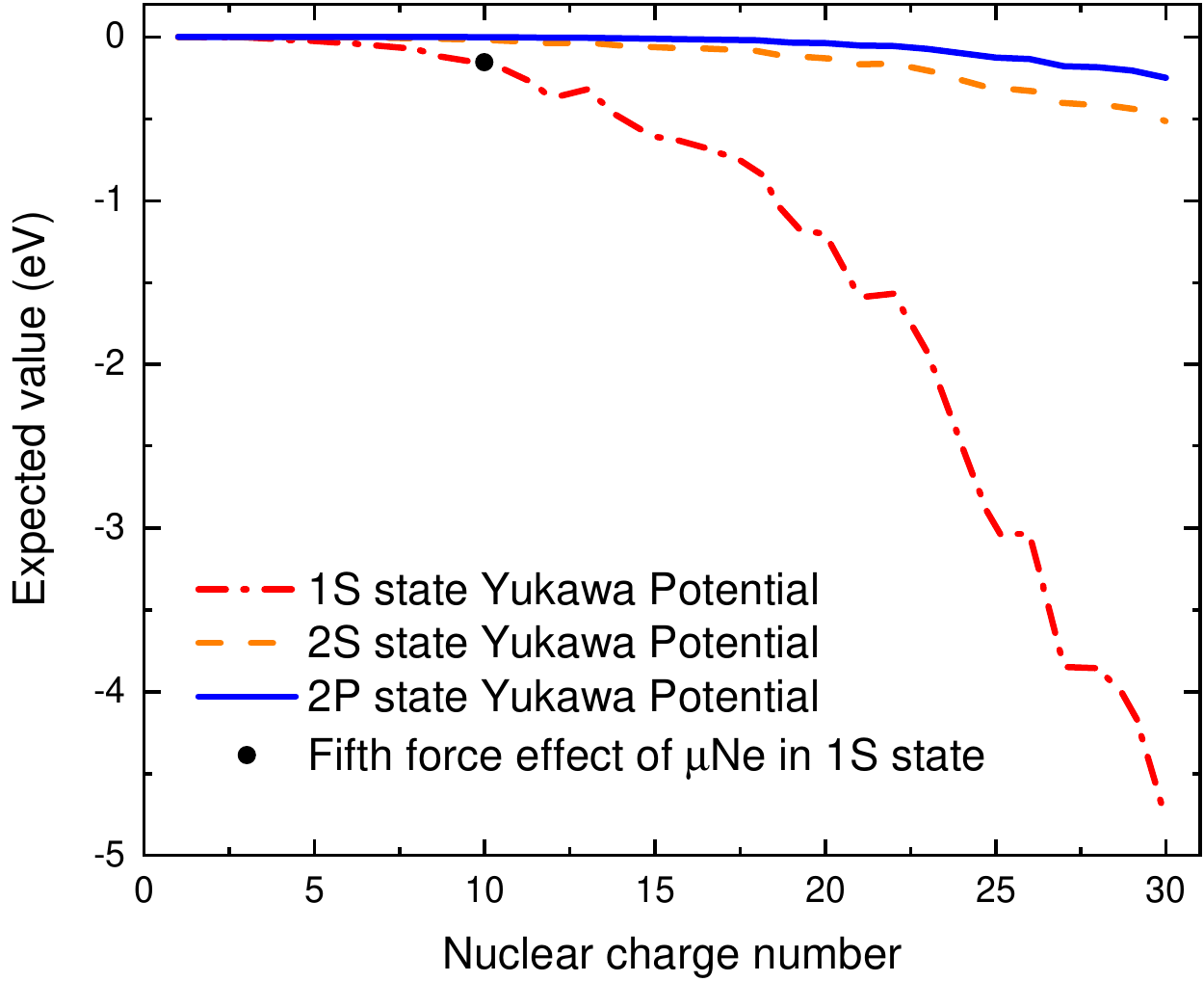}
			\end{minipage}
		}
		\subfigure[]
		{
			\begin{minipage}[htbp]{.45\linewidth}
				\centering
				\includegraphics[scale=0.33]{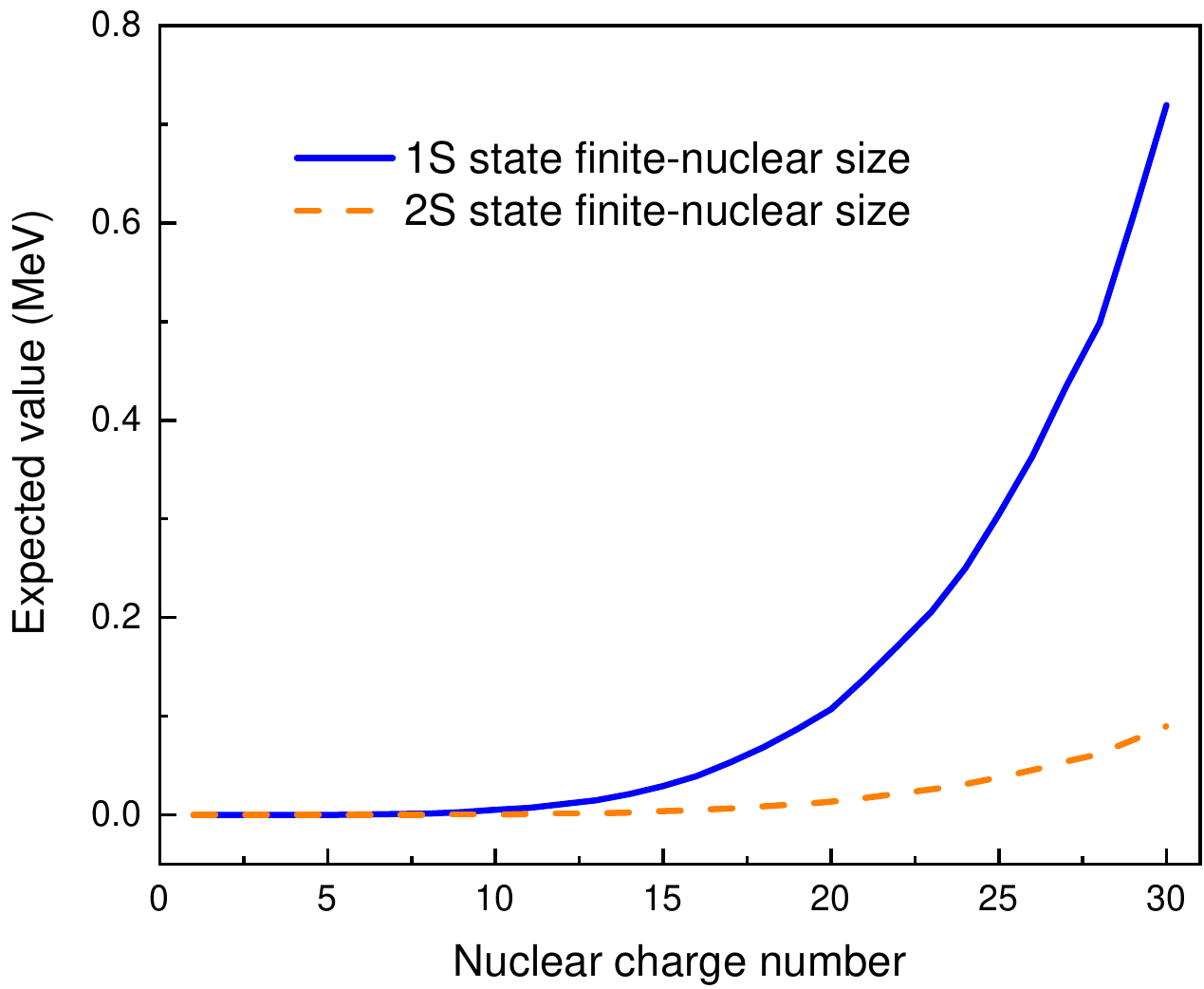}
			\end{minipage}
		}
		\caption{(a) The calculation results of the fifth force. The blue curve represents the expected value of 2P state of the fifth force in muonic systems while the red dashed-dotted curve and the orange dashed curve represent the 1S and 2S states expected values. (b) The calculation results of the finite-nuclear size. The blue curve represents the expected value of 1S state of the finite-nuclear-size in muonic systems while the orange dashed curve represents the 2S state expected values. The impact of fifth force increases with the nuclear charge number. At the same time, the effect of finite-nuclear-size also increases rapidly. And the value of finite-nuclear-size effect is approximately $ 10^{5} $ times larger than the effect of fifth force in nS states of two-body system.}
		\label{fig:fns}
	\end{figure*}
	
Now, we aim to enhance the effect of the fifth force on energy levels and undertake calculations for the fifth force in heavy atomic nuclei. As presented in Fig.~\ref{fig:fns}, the effect of the fifth force on 1S, 2S and 2P states and finite-nuclear-size in S states are examined. For S states, the effect of the fifth force should cause the energy level to move down. Especially the $ \mu$Ne atom, the effect of the fifth force on the energy level of its 1S state is around 0.15 eV which achieves the experimental accuracy ~\cite{PhysRevLett.130.173001}. It means that it is promising to detect the fifth force in 1S state of $ \mu$Ne atom. In addition, it is easy to find the absolute value 2P state is smaller than the 1S state from the Fig.~\ref{fig:fns}. As a result, energy level difference would be larger than if fifth force is ignored.

\subsection{The Effect of Hyperfine Structure of the Fifth Force}%3.1
	
\begin{figure*}[htbp]
		\centering
		\subfigure[]
		{
			\begin{minipage}[htbp]{.45\linewidth}
				\centering
				\includegraphics[scale=0.33]{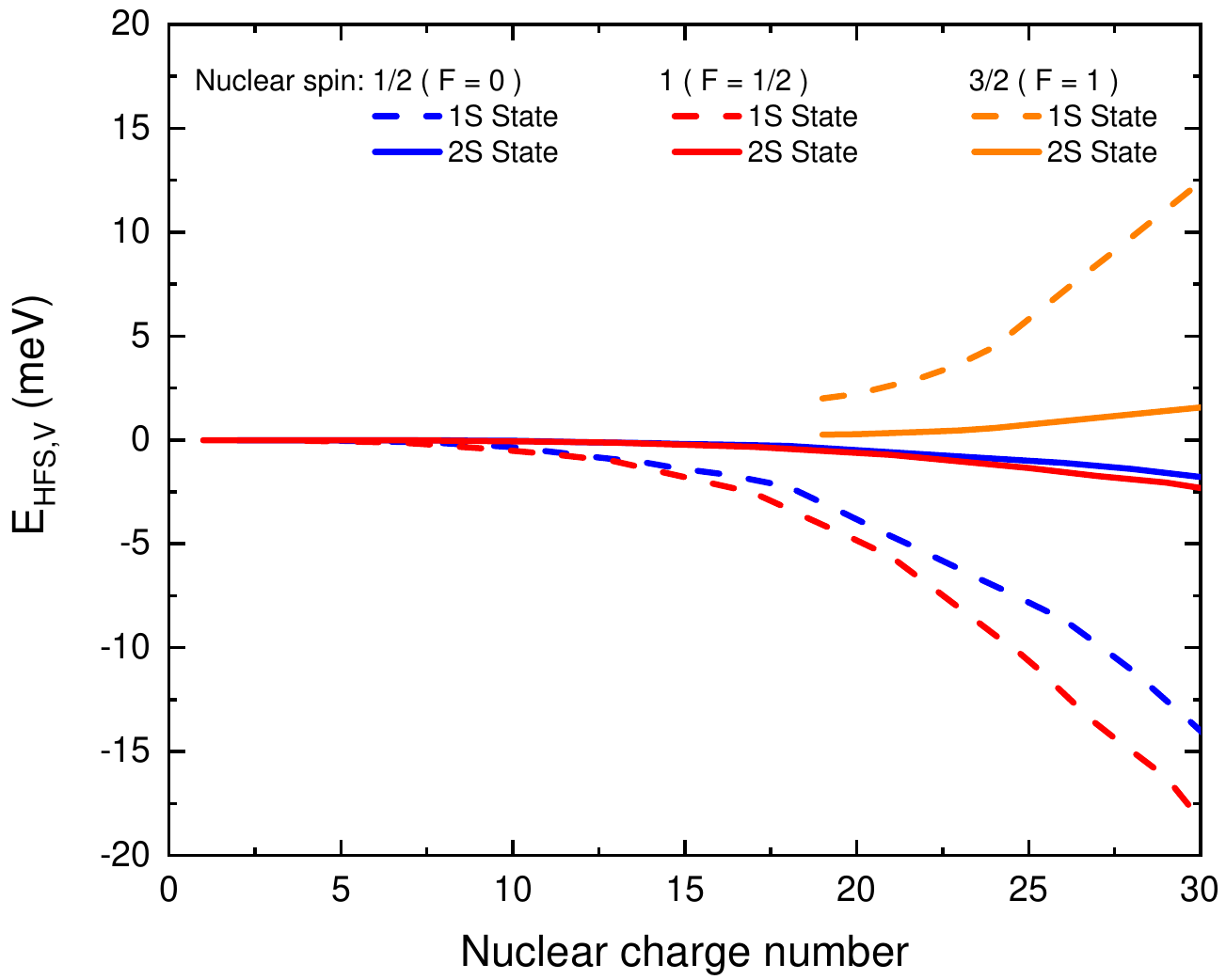}
			\end{minipage}
		}
		\subfigure[]
		{
			\begin{minipage}[htbp]{.45\linewidth}
				\centering
				\includegraphics[scale=0.33]{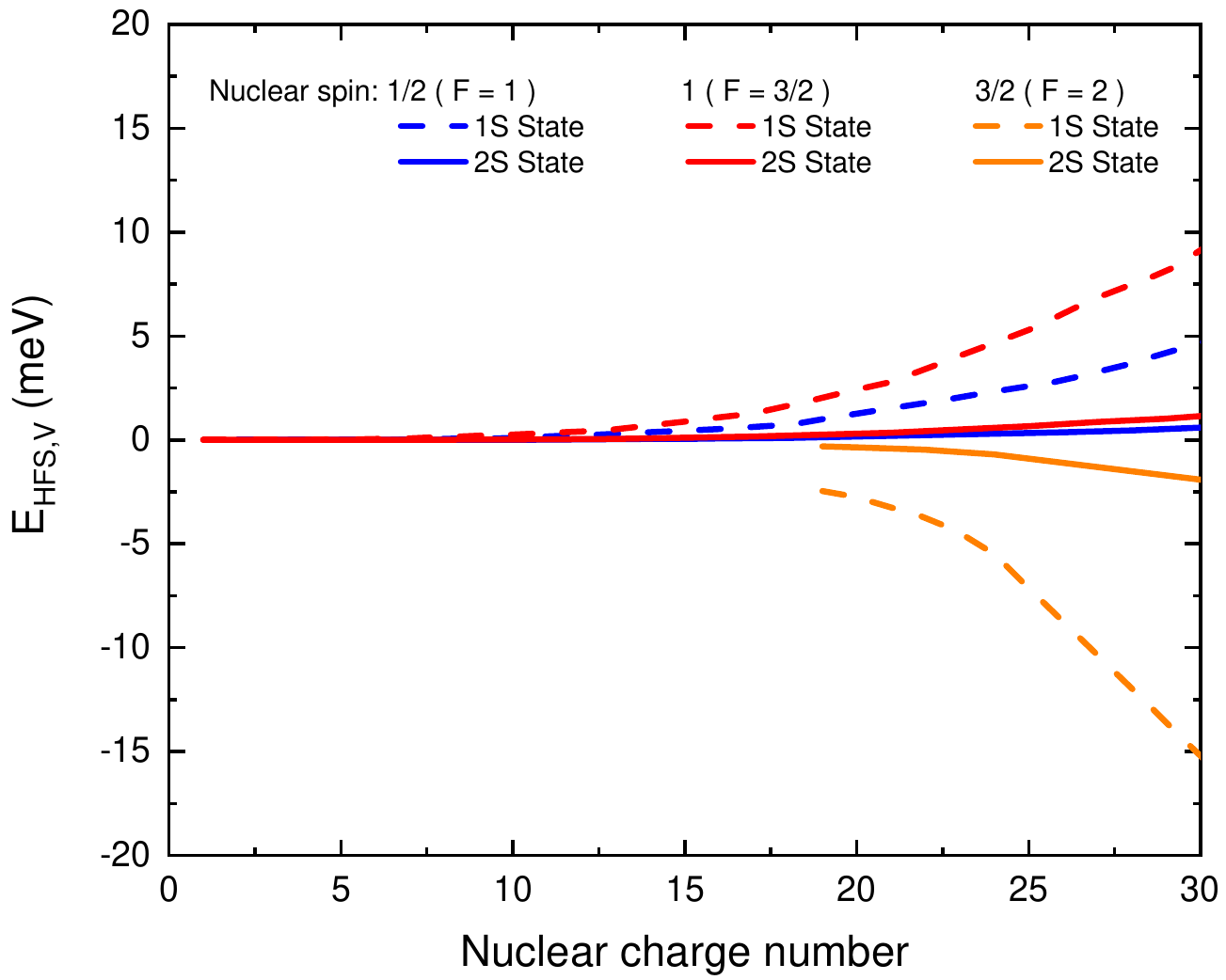}
			\end{minipage}
		}
		
		\subfigure[]
		{
			\begin{minipage}[htbp]{.45\linewidth}
				\centering
				\includegraphics[scale=0.33]{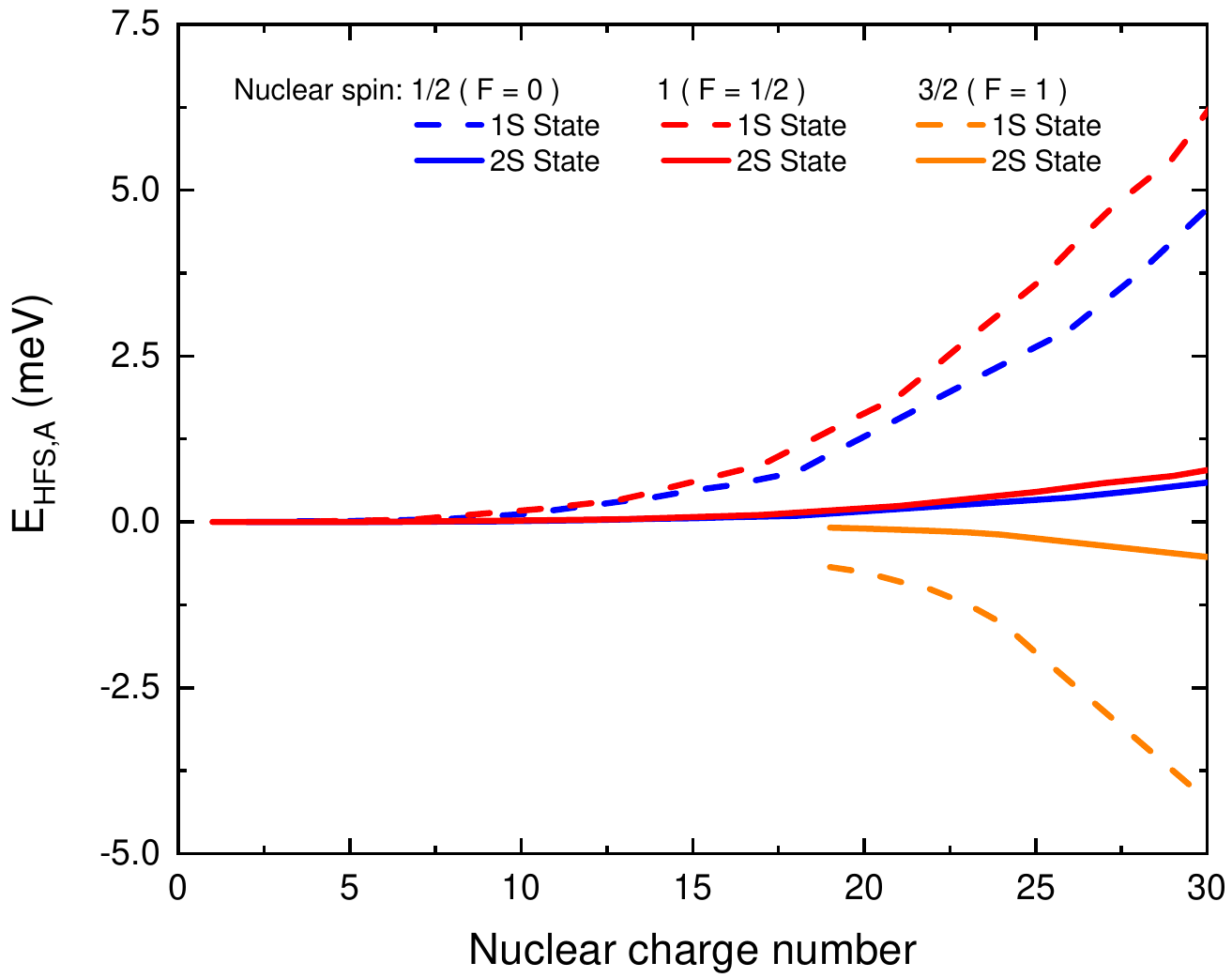}
			\end{minipage}
		}
		\subfigure[]
		{
			\begin{minipage}[htbp]{.45\linewidth}
				\centering
				\includegraphics[scale=0.33]{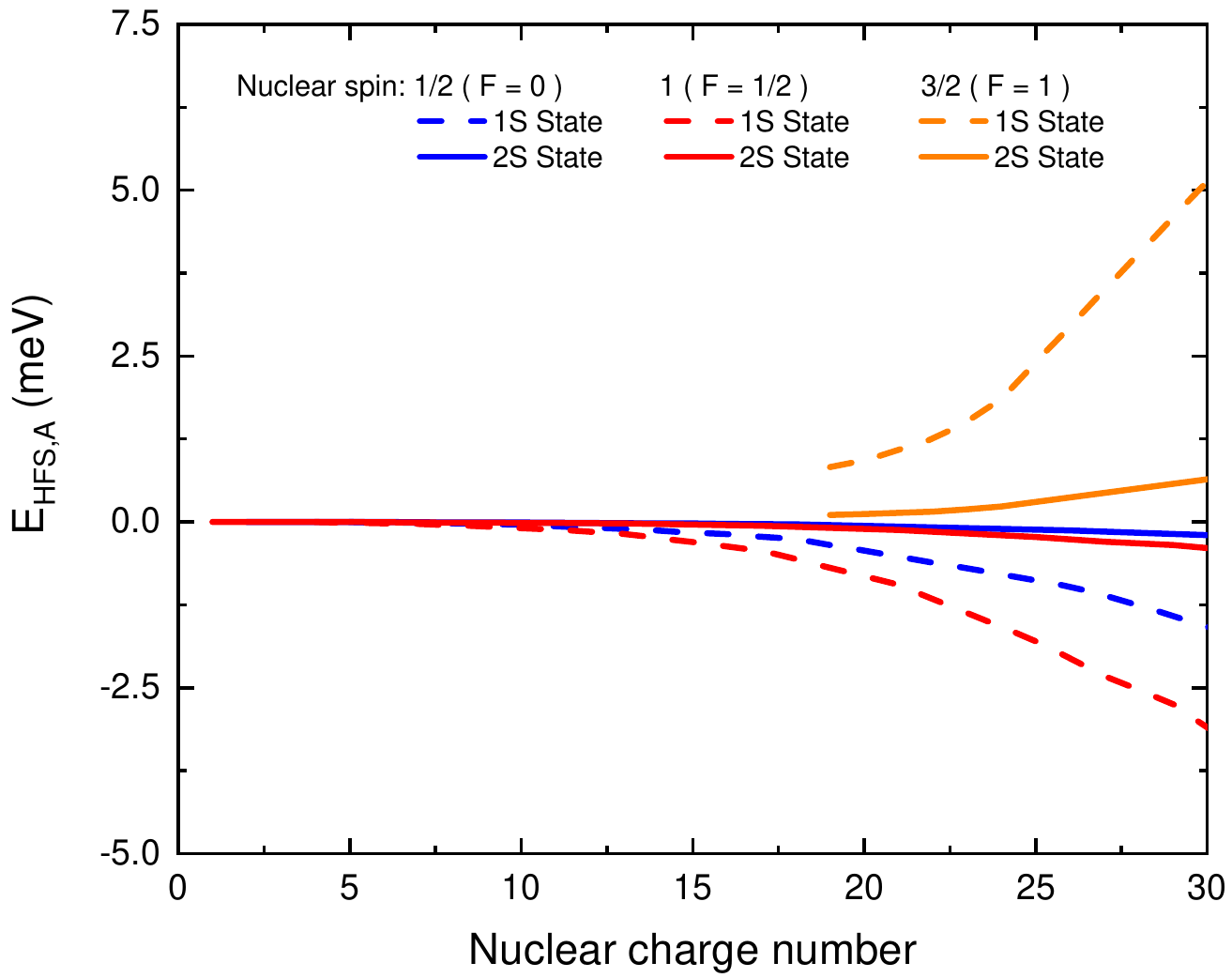}
			\end{minipage}
		}
	    \subfigure[]
	    {
	    	\begin{minipage}[htbp]{.45\linewidth}
	    		\centering
	    		\includegraphics[scale=0.33]{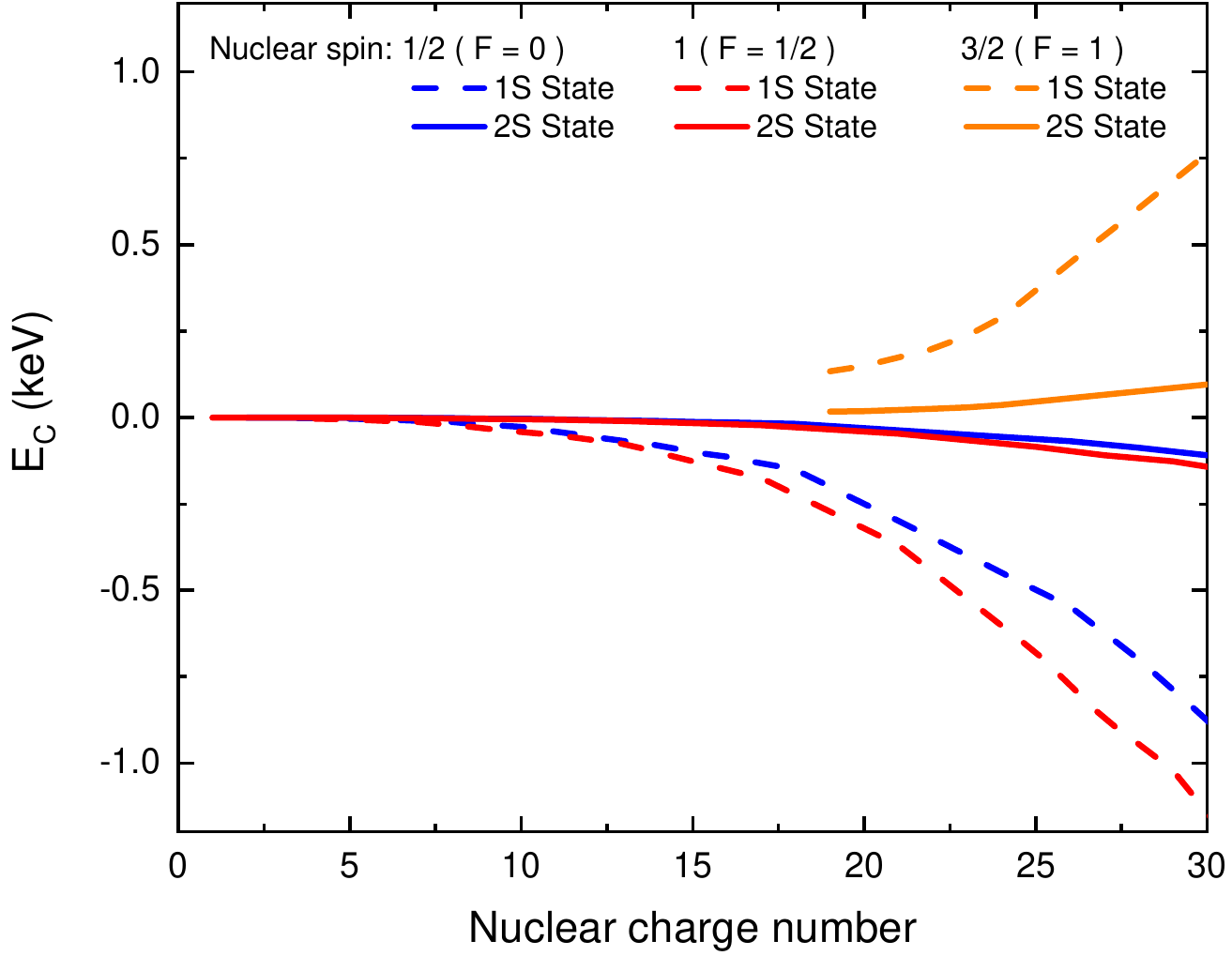}
	    	\end{minipage}
	    }
	    \subfigure[]
	    {
	    	\begin{minipage}[htbp]{.45\linewidth}
	    		\centering
	    		\includegraphics[scale=0.33]{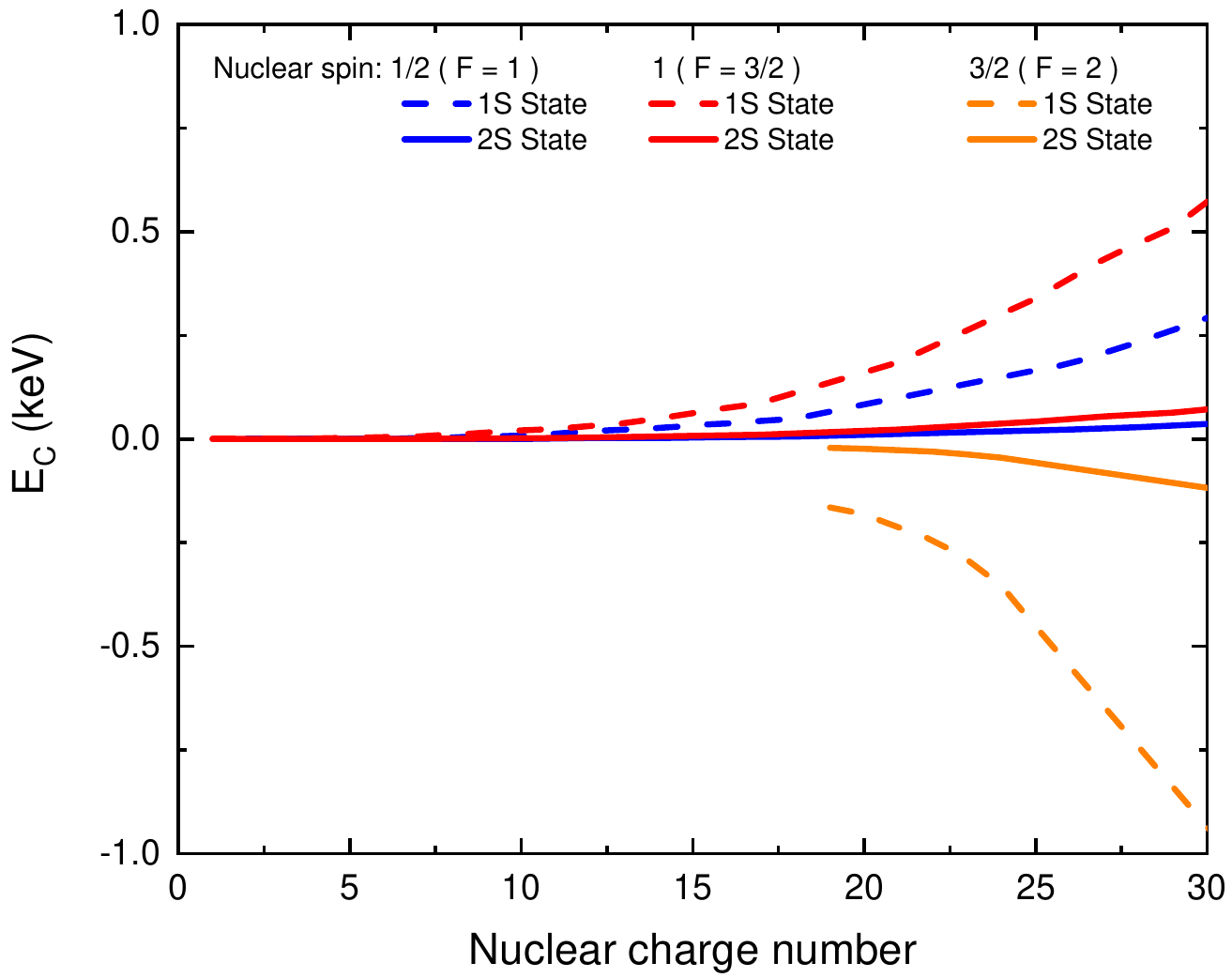}
	    	\end{minipage}
	    }
		\caption{ (Color online) All the results are calculated in the S states. (a) and (b) are the results of the vector particle effect; (c) and (d) are the results of the pseudoscalar particle effect; (e) and (f) are the results of the hypefine structure in electronic systems. Here the solid curves and the dashed curves represent the expected values of the 1S and 2S states,respectively. And different nuclear spin of atoms are distinguished by color: blue represents the nuclear spin of 1/2, red represents the nuclear spin of 1, orange represents the nuclear spin of 3/2. } 
		\label{fig:S-hfs}
	\end{figure*}

\begin{figure*}[htbp]
		\centering
		\subfigure[]
		{
			\begin{minipage}[htbp]{.45\linewidth}
				\centering
				\includegraphics[scale=0.33]{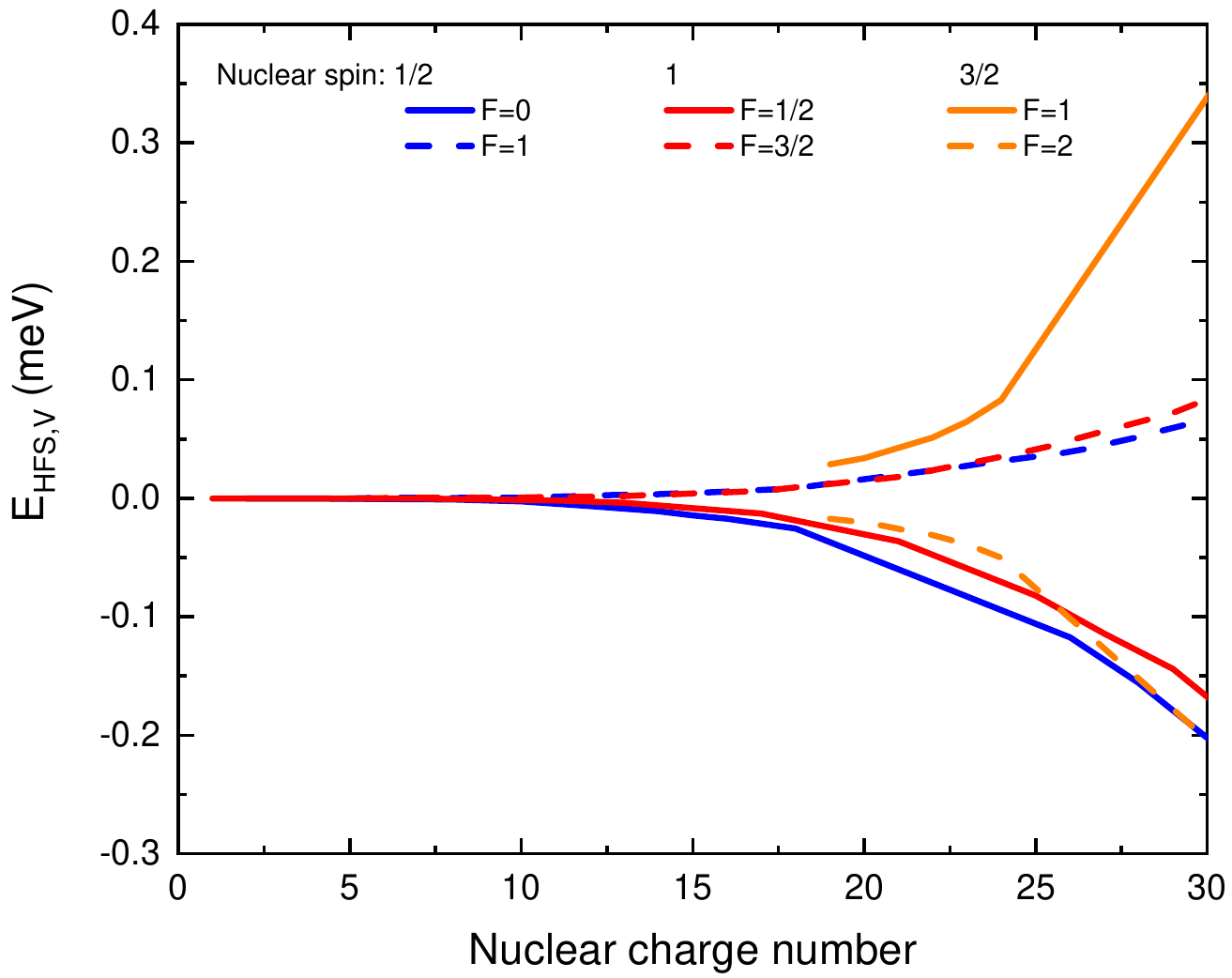}
			\end{minipage}
		}
		\subfigure[]
		{
			\begin{minipage}[htbp]{.45\linewidth}
				\centering
				\includegraphics[scale=0.33]{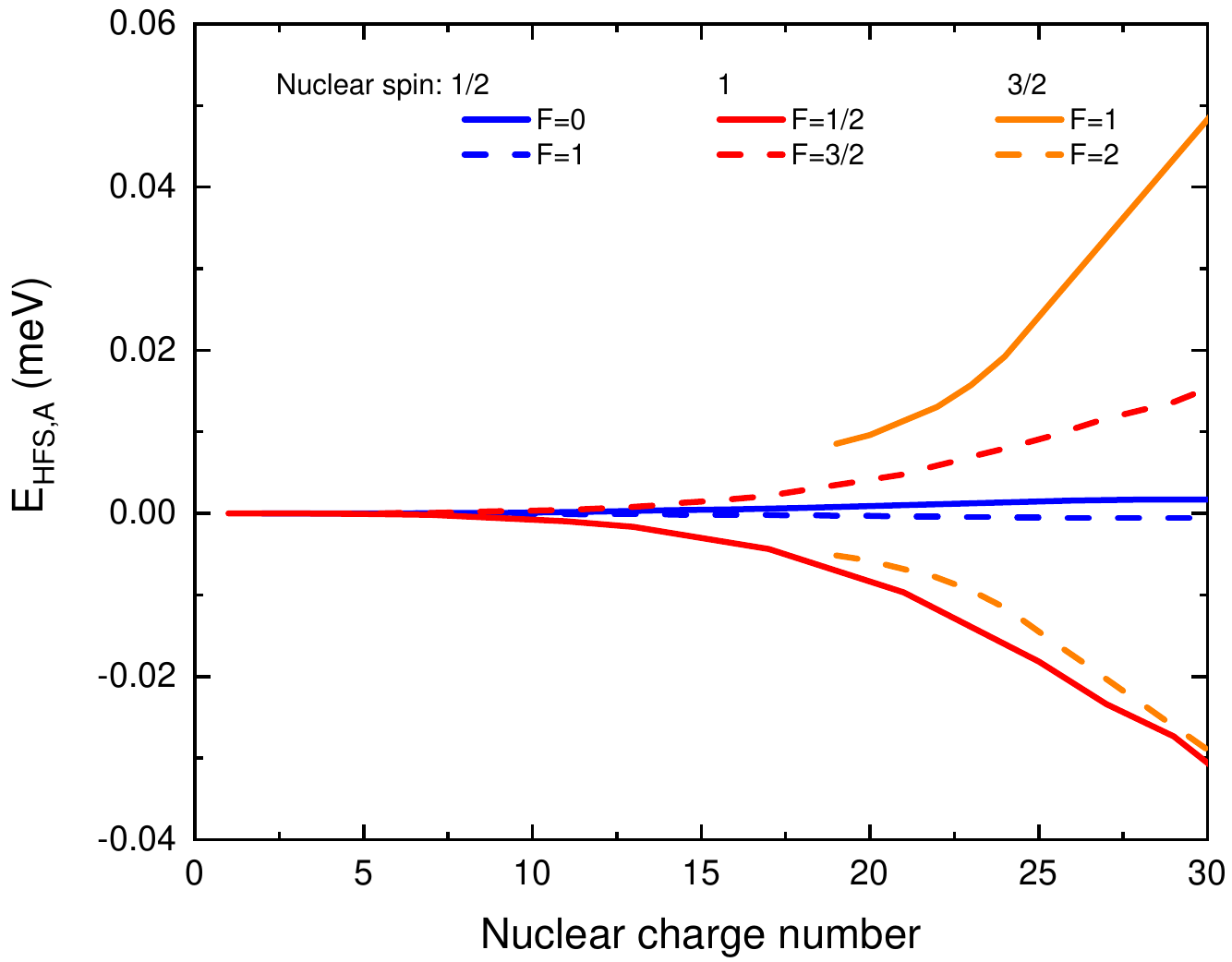}
			\end{minipage}
		}
	    \subfigure[]
	    {
	    	\begin{minipage}[htbp]{.45\linewidth}
	    		\centering
	    		\includegraphics[scale=0.33]{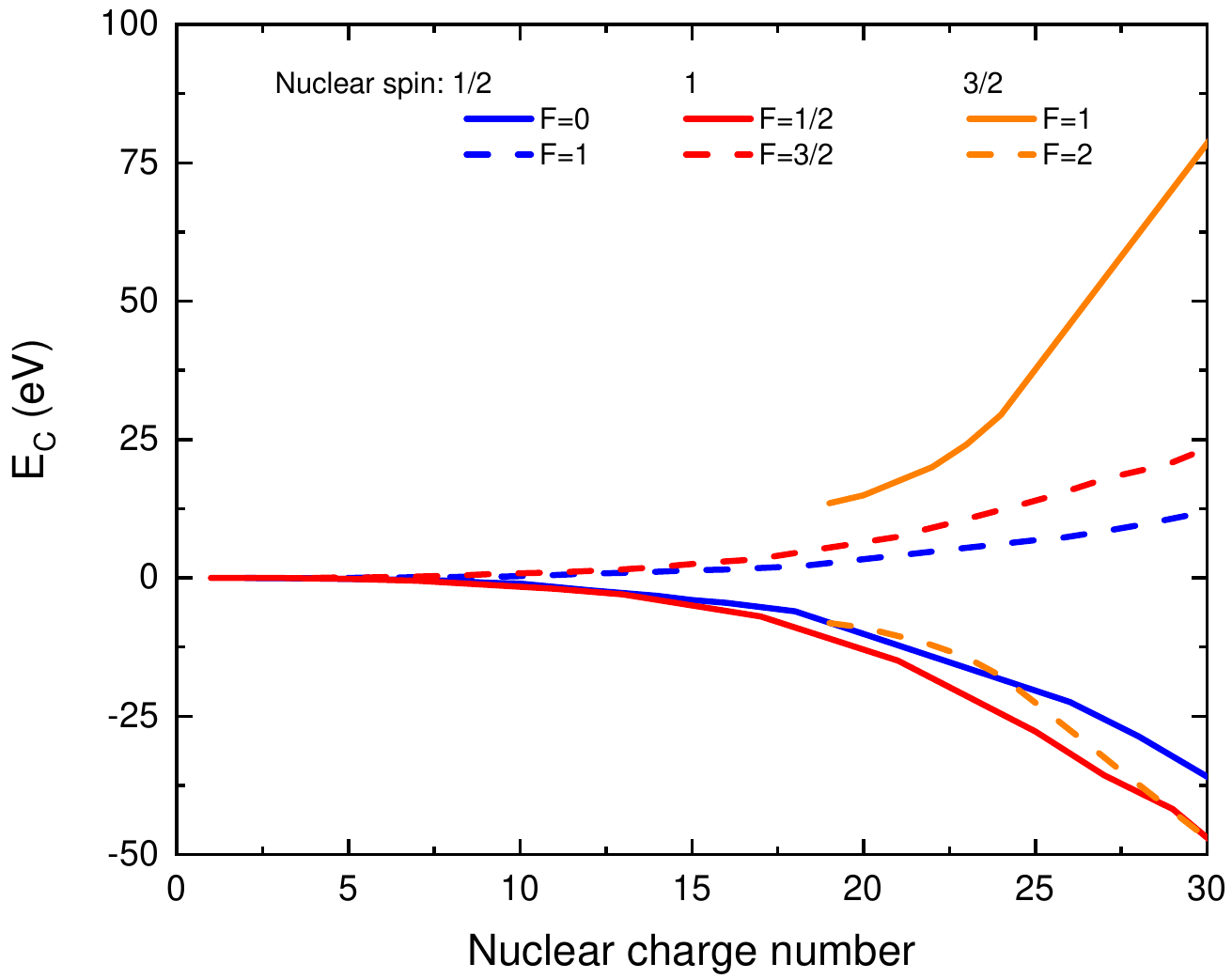}
	    	\end{minipage}
	    }
		\caption{(Color online) All the results are calculated in the P state. (a) The results of the vector particle effect; (b) The results of the pseudoscalar particle effect; (c) The results of the hypefine structure in electronic systems. Here the solid curves and the dashed curves represent the expected values of the different spin of total system. And different nuclear spin of atoms are distinguished by color: blue represents the nuclear spin of 1/2, red represents the nuclear spin of 1, orange represents the nuclear spin of 3/2.  }
		\label{fig:P-hfs}
	\end{figure*}

In this study, we investigate the effect of the hyperfine structure on the fifth force within the 1S, 2S, and 2P states. It is noteworthy that the nuclear spin varies across different atoms and ions, resulting in distinct absolute magnitudes of the fifth force effect. From the Fig.~\ref{fig:S-hfs} and Fig.~\ref{fig:P-hfs}, it is evident that the absolute values of the fifth force effect increases with the nuclear charge number. In the Fig.~\ref{fig:S-hfs} and Fig.~\ref{fig:P-hfs}, the nuclear spins of nuclei with small nuclear charge numbers are mainly concentrated at 1/2 and 1, resulting in the appearance of the orange curve in the latter half of the coordinate axis. Additionally, a comparison between the results of the 2S state and the 2P state reveals that the values within the S states exceed those within the P state. Upon comparing the effects of vector and pseudoscalar particles, it becomes apparent that the impact of the vector particle surpasses that of the pseudoscalar particle. Comparing the (a) and (e) of Fig~\ref{fig:S-hfs}, the value of hypefine structure in electronic systems is approximately $10^5 $ times larger than the value of the X17 as a vector particle in muonic systems while the X17 as a pseudoscalar particle would have a larger discrepancy with it. Furthermore, the same conclusion can be obtained in the P-state from the Fig~\ref{fig:P-hfs}. Consequently, the possibility of detecting the fifth force effect originating from the vector particle is higher. 
	
Although the fifth force is expected to have a greater impact on the energy levels of hyperfine splitting, the magnitude of the hyperfine structure is too small to detect the fifth force experimentally. Additionally, the effects of a vector particle and a pseudoscalar particle on energy levels differ by only a single factor, making the effect of a pseudoscalar particle smaller than that of a vector particle in fine structure. Consequently, the 1S state in fine structure provides the most promising opportunity for detecting the X17 as a vector particle.

\section{Summary}
\label{sec:summary}
	
This study presents a comprehensive analysis of the impact of the potential fifth force on fine and hyperfine splitting, specifically in the 1S state, 2S state, and 2P state. To isolate the effects of the fifth force, we employ a range of approximations, varying between the S and P states, to distinguish between the finite-nuclear-size effect and the fifth force effect. 
 
Based on the accuracy calculation results, it is evident that the absolute value of the fifth force effect increases with the nuclear charge number, regardless of whether it is in the S states or P states. This holds true for both vector and pseudoscalar particles, as well as for both fine structure and hyperfine structure.
	
In a noteworthy experiment reported in Ref.~\cite{PhysRevLett.130.173001}, researchers have achieved a significant milestone by successfully demonstrating the feasibility of strong-field quantum electrodynamics using muonic Neon atoms. Utilizing state-of-the-art X-ray detectors, this proof-of-concept experiment has attained an exceptional level of precision, with a relative error of 0.002$\%$ and a precision of 0.1 eV. This degree of accuracy is in close agreement with the calculated value of the fifth force effect in muonic atoms with nuclear charge number ranging from 10 to 15, which heightens the prospect of discovering this elusive phenomenon.

\begin{acknowledgments}
This work is supported by the Strategic Priority Research Program of Chinese Academy of Sciences (Grant No. XDB34030301), the National Natural Science Foundation of China with Grant No.~12375073 and Major Project of Basic and Applied Basic Research in Guangdong Province 2020B0301030008.	
\end{acknowledgments}

\bibliographystyle{apsrev4-1}
\bibliography{fifth_force}

\begin{thebibliography}{16}%
\makeatletter
\providecommand \@ifxundefined [1]{%
 \@ifx{#1\undefined}
}%
\providecommand \@ifnum [1]{%
 \ifnum #1\expandafter \@firstoftwo
 \else \expandafter \@secondoftwo
 \fi
}%
\providecommand \@ifx [1]{%
 \ifx #1\expandafter \@firstoftwo
 \else \expandafter \@secondoftwo
 \fi
}%
\providecommand \natexlab [1]{#1}%
\providecommand \enquote  [1]{``#1''}%
\providecommand \bibnamefont  [1]{#1}%
\providecommand \bibfnamefont [1]{#1}%
\providecommand \citenamefont [1]{#1}%
\providecommand \href@noop [0]{\@secondoftwo}%
\providecommand \href [0]{\begingroup \@sanitize@url \@href}%
\providecommand \@href[1]{\@@startlink{#1}\@@href}%
\providecommand \@@href[1]{\endgroup#1\@@endlink}%
\providecommand \@sanitize@url [0]{\catcode `\\12\catcode `\$12\catcode `\&12\catcode `\#12\catcode `\^12\catcode `\_12\catcode `\%12\relax}%
\providecommand \@@startlink[1]{}%
\providecommand \@@endlink[0]{}%
\providecommand \url  [0]{\begingroup\@sanitize@url \@url }%
\providecommand \@url [1]{\endgroup\@href {#1}{\urlprefix }}%
\providecommand \urlprefix  [0]{URL }%
\providecommand \Eprint [0]{\href }%
\providecommand \doibase [0]{http://dx.doi.org/}%
\providecommand \selectlanguage [0]{\@gobble}%
\providecommand \bibinfo  [0]{\@secondoftwo}%
\providecommand \bibfield  [0]{\@secondoftwo}%
\providecommand \translation [1]{[#1]}%
\providecommand \BibitemOpen [0]{}%
\providecommand \bibitemStop [0]{}%
\providecommand \bibitemNoStop [0]{.\EOS\space}%
\providecommand \EOS [0]{\spacefactor3000\relax}%
\providecommand \BibitemShut  [1]{\csname bibitem#1\endcsname}%
\let\auto@bib@innerbib\@empty
%</preamble>
\bibitem [{\citenamefont {Sushkov}\ \emph {et~al.}(2011)\citenamefont {Sushkov}, \citenamefont {Kim}, \citenamefont {Dalvit},\ and\ \citenamefont {Lamoreaux}}]{Sushkov:2011md}%
  \BibitemOpen
  \bibfield  {author} {\bibinfo {author} {\bibfnamefont {A.~O.}\ \bibnamefont {Sushkov}}, \bibinfo {author} {\bibfnamefont {W.~J.}\ \bibnamefont {Kim}}, \bibinfo {author} {\bibfnamefont {D.~A.~R.}\ \bibnamefont {Dalvit}}, \ and\ \bibinfo {author} {\bibfnamefont {S.~K.}\ \bibnamefont {Lamoreaux}},\ }\href {\doibase 10.1103/PhysRevLett.107.171101} {\bibfield  {journal} {\bibinfo  {journal} {Phys. Rev. Lett.}\ }\textbf {\bibinfo {volume} {107}},\ \bibinfo {pages} {171101} (\bibinfo {year} {2011})},\ \Eprint {http://arxiv.org/abs/1108.2547} {arXiv:1108.2547 [quant-ph]} \BibitemShut {NoStop}%
\bibitem [{\citenamefont {Krasznahorkay}\ \emph {et~al.}(2016)\citenamefont {Krasznahorkay} \emph {et~al.}}]{Krasznahorkay:2015iga}%
  \BibitemOpen
  \bibfield  {author} {\bibinfo {author} {\bibfnamefont {A.~J.}\ \bibnamefont {Krasznahorkay}} \emph {et~al.},\ }\href {\doibase 10.1103/PhysRevLett.116.042501} {\bibfield  {journal} {\bibinfo  {journal} {Phys. Rev. Lett.}\ }\textbf {\bibinfo {volume} {116}},\ \bibinfo {pages} {042501} (\bibinfo {year} {2016})},\ \Eprint {http://arxiv.org/abs/1504.01527} {arXiv:1504.01527 [nucl-ex]} \BibitemShut {NoStop}%
\bibitem [{\citenamefont {Krasznahorkay}\ \emph {et~al.}(2017)\citenamefont {Krasznahorkay} \emph {et~al.}}]{Krasznahorkay:2017gwn}%
  \BibitemOpen
  \bibfield  {author} {\bibinfo {author} {\bibfnamefont {A.~J.}\ \bibnamefont {Krasznahorkay}} \emph {et~al.},\ }\href {\doibase 10.1051/epjconf/201714201019} {\bibfield  {journal} {\bibinfo  {journal} {EPJ Web Conf.}\ }\textbf {\bibinfo {volume} {142}},\ \bibinfo {pages} {01019} (\bibinfo {year} {2017})}\BibitemShut {NoStop}%
\bibitem [{\citenamefont {Krasznahorkay}\ \emph {et~al.}(2019)\citenamefont {Krasznahorkay} \emph {et~al.}}]{Krasznahorkay:2019lyl}%
  \BibitemOpen
  \bibfield  {author} {\bibinfo {author} {\bibfnamefont {A.~J.}\ \bibnamefont {Krasznahorkay}} \emph {et~al.},\ }\href@noop {} {\  (\bibinfo {year} {2019})},\ \Eprint {http://arxiv.org/abs/1910.10459} {arXiv:1910.10459 [nucl-ex]} \BibitemShut {NoStop}%
\bibitem [{\citenamefont {Feng}\ \emph {et~al.}(2016)\citenamefont {Feng}, \citenamefont {Fornal}, \citenamefont {Galon}, \citenamefont {Gardner}, \citenamefont {Smolinsky}, \citenamefont {Tait},\ and\ \citenamefont {Tanedo}}]{Feng:2016jff}%
  \BibitemOpen
  \bibfield  {author} {\bibinfo {author} {\bibfnamefont {J.~L.}\ \bibnamefont {Feng}}, \bibinfo {author} {\bibfnamefont {B.}~\bibnamefont {Fornal}}, \bibinfo {author} {\bibfnamefont {I.}~\bibnamefont {Galon}}, \bibinfo {author} {\bibfnamefont {S.}~\bibnamefont {Gardner}}, \bibinfo {author} {\bibfnamefont {J.}~\bibnamefont {Smolinsky}}, \bibinfo {author} {\bibfnamefont {T.~M.~P.}\ \bibnamefont {Tait}}, \ and\ \bibinfo {author} {\bibfnamefont {P.}~\bibnamefont {Tanedo}},\ }\href {\doibase 10.1103/PhysRevLett.117.071803} {\bibfield  {journal} {\bibinfo  {journal} {Phys. Rev. Lett.}\ }\textbf {\bibinfo {volume} {117}},\ \bibinfo {pages} {071803} (\bibinfo {year} {2016})},\ \Eprint {http://arxiv.org/abs/1604.07411} {arXiv:1604.07411 [hep-ph]} \BibitemShut {NoStop}%
\bibitem [{\citenamefont {Feng}\ \emph {et~al.}(2017)\citenamefont {Feng}, \citenamefont {Fornal}, \citenamefont {Galon}, \citenamefont {Gardner}, \citenamefont {Smolinsky}, \citenamefont {Tait},\ and\ \citenamefont {Tanedo}}]{Feng:2016ysn}%
  \BibitemOpen
  \bibfield  {author} {\bibinfo {author} {\bibfnamefont {J.~L.}\ \bibnamefont {Feng}}, \bibinfo {author} {\bibfnamefont {B.}~\bibnamefont {Fornal}}, \bibinfo {author} {\bibfnamefont {I.}~\bibnamefont {Galon}}, \bibinfo {author} {\bibfnamefont {S.}~\bibnamefont {Gardner}}, \bibinfo {author} {\bibfnamefont {J.}~\bibnamefont {Smolinsky}}, \bibinfo {author} {\bibfnamefont {T.~M.~P.}\ \bibnamefont {Tait}}, \ and\ \bibinfo {author} {\bibfnamefont {P.}~\bibnamefont {Tanedo}},\ }\href {\doibase 10.1103/PhysRevD.95.035017} {\bibfield  {journal} {\bibinfo  {journal} {Phys. Rev. D}\ }\textbf {\bibinfo {volume} {95}},\ \bibinfo {pages} {035017} (\bibinfo {year} {2017})},\ \Eprint {http://arxiv.org/abs/1608.03591} {arXiv:1608.03591 [hep-ph]} \BibitemShut {NoStop}%
\bibitem [{\citenamefont {Ellwanger}\ and\ \citenamefont {Moretti}(2016)}]{Ellwanger:2016wfe}%
  \BibitemOpen
  \bibfield  {author} {\bibinfo {author} {\bibfnamefont {U.}~\bibnamefont {Ellwanger}}\ and\ \bibinfo {author} {\bibfnamefont {S.}~\bibnamefont {Moretti}},\ }\href {\doibase 10.1007/JHEP11(2016)039} {\bibfield  {journal} {\bibinfo  {journal} {JHEP}\ }\textbf {\bibinfo {volume} {11}},\ \bibinfo {pages} {039} (\bibinfo {year} {2016})},\ \Eprint {http://arxiv.org/abs/1609.01669} {arXiv:1609.01669 [hep-ph]} \BibitemShut {NoStop}%
\bibitem [{\citenamefont {Jentschura}\ and\ \citenamefont {N\'andori}(2018)}]{Jentschura:2018zjv}%
  \BibitemOpen
  \bibfield  {author} {\bibinfo {author} {\bibfnamefont {U.~D.}\ \bibnamefont {Jentschura}}\ and\ \bibinfo {author} {\bibfnamefont {I.}~\bibnamefont {N\'andori}},\ }\href {\doibase 10.1103/PhysRevA.97.042502} {\bibfield  {journal} {\bibinfo  {journal} {Phys. Rev. A}\ }\textbf {\bibinfo {volume} {97}},\ \bibinfo {pages} {042502} (\bibinfo {year} {2018})},\ \Eprint {http://arxiv.org/abs/1804.03096} {arXiv:1804.03096 [hep-ph]} \BibitemShut {NoStop}%
\bibitem [{\citenamefont {Jentschura}(2020)}]{Jentschura:2020zlr}%
  \BibitemOpen
  \bibfield  {author} {\bibinfo {author} {\bibfnamefont {U.~D.}\ \bibnamefont {Jentschura}},\ }\href {\doibase 10.1103/PhysRevA.101.062503} {\bibfield  {journal} {\bibinfo  {journal} {Phys. Rev. A}\ }\textbf {\bibinfo {volume} {101}},\ \bibinfo {pages} {062503} (\bibinfo {year} {2020})},\ \Eprint {http://arxiv.org/abs/2003.07207} {arXiv:2003.07207 [hep-ph]} \BibitemShut {NoStop}%
\bibitem [{\citenamefont {Sapirstein}\ and\ \citenamefont {Yennie}(1990)}]{1990THEORY}%
  \BibitemOpen
  \bibfield  {author} {\bibinfo {author} {\bibfnamefont {J.~R.}\ \bibnamefont {Sapirstein}}\ and\ \bibinfo {author} {\bibfnamefont {D.~R.}\ \bibnamefont {Yennie}},\ }\href@noop {} {\emph {\bibinfo {title} {THEORY OF HYDROGENIC BOUND STATES}}}\ (\bibinfo  {publisher} {Quantum Electrodynamics},\ \bibinfo {year} {1990})\BibitemShut {NoStop}%
\bibitem [{\citenamefont {Jentschura}\ and\ \citenamefont {Pachucki}(1996)}]{Jentschura:1996zz}%
  \BibitemOpen
  \bibfield  {author} {\bibinfo {author} {\bibfnamefont {U.}~\bibnamefont {Jentschura}}\ and\ \bibinfo {author} {\bibfnamefont {K.}~\bibnamefont {Pachucki}},\ }\href {\doibase 10.1103/PhysRevA.54.1853} {\bibfield  {journal} {\bibinfo  {journal} {Phys. Rev. A}\ }\textbf {\bibinfo {volume} {54}},\ \bibinfo {pages} {1853} (\bibinfo {year} {1996})},\ \Eprint {http://arxiv.org/abs/physics/0011008} {arXiv:physics/0011008} \BibitemShut {NoStop}%
\bibitem [{\citenamefont {Hiyama}\ \emph {et~al.}(2003)\citenamefont {Hiyama}, \citenamefont {Kino},\ and\ \citenamefont {Kamimura}}]{Hiyama2003}%
  \BibitemOpen
  \bibfield  {author} {\bibinfo {author} {\bibfnamefont {E.}~\bibnamefont {Hiyama}}, \bibinfo {author} {\bibfnamefont {Y.}~\bibnamefont {Kino}}, \ and\ \bibinfo {author} {\bibfnamefont {M.}~\bibnamefont {Kamimura}},\ }\href {\doibase 10.1016/S0146-6410(03)90015-9} {\bibfield  {journal} {\bibinfo  {journal} {Prog. Part. Nucl. Phys.}\ }\textbf {\bibinfo {volume} {51}},\ \bibinfo {pages} {223} (\bibinfo {year} {2003})}\BibitemShut {NoStop}%
\bibitem [{\citenamefont {et~al.}(2023)}]{PhysRevLett.130.173001}%
  \BibitemOpen
  \bibfield  {author} {\bibinfo {author} {\bibfnamefont {T.~O.}\ \bibnamefont {et~al.}},\ }\href {\doibase 10.1103/PhysRevLett.130.173001} {\bibfield  {journal} {\bibinfo  {journal} {Phys. Rev. Lett.}\ }\textbf {\bibinfo {volume} {130}},\ \bibinfo {pages} {173001} (\bibinfo {year} {2023})}\BibitemShut {NoStop}%
\bibitem [{\citenamefont {Angeli}(2004)}]{Angeli:2004kvy}%
  \BibitemOpen
  \bibfield  {author} {\bibinfo {author} {\bibfnamefont {I.}~\bibnamefont {Angeli}},\ }\href {\doibase 10.1016/j.adt.2004.04.002} {\bibfield  {journal} {\bibinfo  {journal} {Atom. Data Nucl. Data Tabl.}\ }\textbf {\bibinfo {volume} {87}},\ \bibinfo {pages} {185} (\bibinfo {year} {2004})}\BibitemShut {NoStop}%
\bibitem [{\citenamefont {Jentschura}\ and\ \citenamefont {Yerokhin}(2006)}]{PhysRevA.73.062503}%
  \BibitemOpen
  \bibfield  {author} {\bibinfo {author} {\bibfnamefont {U.~D.}\ \bibnamefont {Jentschura}}\ and\ \bibinfo {author} {\bibfnamefont {V.~A.}\ \bibnamefont {Yerokhin}},\ }\href {\doibase 10.1103/PhysRevA.73.062503} {\bibfield  {journal} {\bibinfo  {journal} {Phys. Rev. A}\ }\textbf {\bibinfo {volume} {73}},\ \bibinfo {pages} {062503} (\bibinfo {year} {2006})}\BibitemShut {NoStop}%
\bibitem [{\citenamefont {Cheng}\ and\ \citenamefont {Chiang}(2012)}]{Cheng:2012qr}%
  \BibitemOpen
  \bibfield  {author} {\bibinfo {author} {\bibfnamefont {H.-Y.}\ \bibnamefont {Cheng}}\ and\ \bibinfo {author} {\bibfnamefont {C.-W.}\ \bibnamefont {Chiang}},\ }\href {\doibase 10.1007/JHEP07(2012)009} {\bibfield  {journal} {\bibinfo  {journal} {JHEP}\ }\textbf {\bibinfo {volume} {07}},\ \bibinfo {pages} {009} (\bibinfo {year} {2012})},\ \Eprint {http://arxiv.org/abs/1202.1292} {arXiv:1202.1292 [hep-ph]} \BibitemShut {NoStop}%
\end{thebibliography}%

\end{document}